\documentclass[aps,prl,amsmath,amssymb,reprint,superscriptaddress]{revtex4-1}

\usepackage[utf8]{inputenc}
\usepackage[T1]{fontenc}
\usepackage{graphicx}
\usepackage[dvipsnames]{xcolor}
\usepackage{dcolumn}
\usepackage{bm}
\usepackage[colorlinks=true, linkcolor=RoyalBlue, citecolor=RoyalBlue]{hyperref}
\usepackage[per-mode=symbol, separate-uncertainty]{siunitx}
\usepackage[capitalise]{cleveref}
\usepackage{enumerate}
\usepackage{float}
\usepackage{lineno}

\begin{document}
\title{Spin-correlated exciton-polaritons in a van der Waals magnet} 
	
\author{Florian Dirnberger}
\email{fdirnberger@ccny.cuny.edu}
\thanks{These authors contributed equally.}
\affiliation{Department of Physics, City College of New York, New York, NY 10031, USA}

\author{Rezlind Bushati}
\thanks{These authors contributed equally.}
\affiliation{Department of Physics, City College of New York, New York, NY 10031, USA}
\affiliation{Department of Physics, The Graduate Center, City University of New York, New York, NY 10016, USA}

\author{Biswajit Datta}
\affiliation{Department of Physics, City College of New York, New York, NY 10031, USA}

\author{Ajesh Kumar}
\affiliation{Department of Physics, University of Texas at Austin, Austin, TX 78712, USA}

\author{Allan H. MacDonald}
\affiliation{Department of Physics, University of Texas at Austin, Austin, TX 78712, USA}

\author{Edoardo Baldini}
\email{edoardo.baldini@austin.utexas.edu}
\affiliation{Department of Physics, University of Texas at Austin, Austin, TX 78712, USA}

\author{Vinod M. Menon}
\email{vmenon@ccny.cuny.edu}
\affiliation{Department of Physics, City College of New York, New York, NY 10031, USA}
\affiliation{Department of Physics, The Graduate Center, City University of New York, New York, NY 10016, USA}


\begin{abstract}
Strong coupling between light and elementary excitations is emerging as a powerful tool to engineer the properties of solid-state systems.
Spin-correlated excitations that couple strongly to optical cavities promise control over collective quantum phenomena such as magnetic phase transitions, but their suitable electronic resonances have yet to be found.
Here we report strong light-matter coupling in $\textrm{NiPS}_3$, a van der Waals antiferromagnet with highly correlated electronic degrees of freedom. 
A previously unobserved class of polaritonic quasiparticles emerges from the strong coupling between its spin-correlated excitons and the photons inside a microcavity.
Detailed spectroscopic analysis in conjunction with a microscopic theory provides unique insights into the origin and interactions of these exotic magnetically coupled excitations.
Our work introduces van der Waals magnets to the field of strong light-matter physics and provides a path towards the design and control of correlated electron systems via cavity quantum electrodynamics.
\end{abstract}

\maketitle
	

Strong light-matter coupling has recently emerged as an attractive platform to engineer quantum materials. 
When coherent optical feedback drives material excitations into a strong coupling regime, new hybrid quasiparticles known as polaritons determine the energy scales of the coupled system. 
Dressing specific collective modes, such as excitons or phonons with tailored optical fields is expected to cause selective modifications of material properties~\cite{Garcia2021,Basov2016}.
Pioneering experiments using excitons in semiconductors embedded in optical cavities realized Bose–Einstein-like condensates, demonstrated fundamental phenomena, such as superfluidity, and spawned device concepts like polaritonic interferometers and Hamiltonian simulators~\cite{Deng2010,Sanvitto2016}.
More recently, the prospects of strong light-matter coupling for manipulating collective phenomena like magnetism, superconductivity, and ferroelectricity in correlated materials has garnered much attention~\cite{Sentef2018,Ashida2020,Thomas2021}.

Among different classes of quantum materials, magnetic van der Waals (vdW) crystals provide access to a large variety of electronic and magnetic phases.
Their strong electronic correlations, exotic magnetic orders, and potential for pressure-induced superconductivity currently motivates immense research efforts~\cite{Gong2019}. 
In addition to magnetic ordering, experiments demonstrated strong internal coupling of phonons and magnons~\cite{Liu2021}.
Such intertwined degrees of freedom make these materials an ideal platform for exploring new aspects of tailored optical control. 
Particularly the recent discovery of optically active excitons with coupling to the antiferromagnetic order~\cite{Kang2020,Hwangbo2021,Wang2021,Belvin2021} opens an extraordinary opportunity in this endeavor. 

Here we demonstrate strong coupling between an optical microcavity mode and spin-correlated excitons hosted in the vdW magnetic insulator $\textrm{NiPS}_3$. 
Hybridization results in a previously unobserved class of polaritons with unique signatures of excitons, photons, and spins.
These newly formed quasiparticles are utilized as a probe to study the nature and interactions of spin-correlated excitons.
We find that long-range excitonic interactions are severely suppressed owing to the tightly bound and highly localized nature of the excitons. 
By modeling excitonic transitions in strongly correlated insulators, we capture the key experimental signatures and demonstrate that spin-correlated excitons in $\textrm{NiPS}_3$ have an origin that is distinct from that of excitons in conventional band semiconductors. 

\begin{figure*}
	\includegraphics[width=0.8\linewidth]{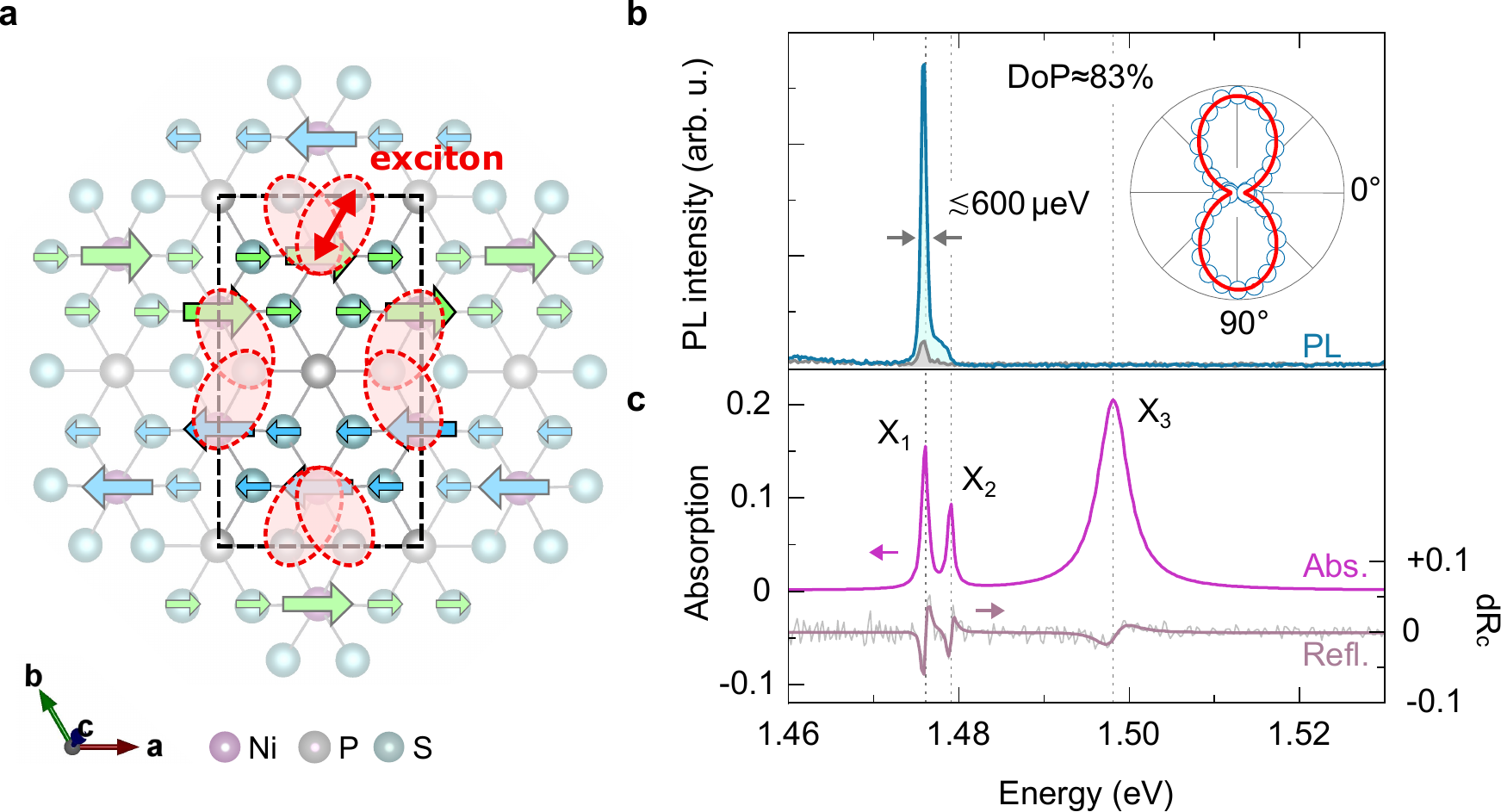}
	\caption{\textbf{Excitons in the vdW antiferromagnet $\textrm{NiPS}_3$.} 
		\textbf{a}~Magnetic moments of Ni atoms in the hexagonal honeycomb lattice slightly magnetize surrounding S ligands except those located between two Ni atoms with opposite spin. Excitons with highly anisotropic dipole moment (red ellipses) form between spin-polarized Ni $d$-orbitals and unmagnetized S $p$-orbitals. The black dashed rectangle marks the in-plane magnetic unit cell with dimensions $\SI{5.8}{\angstrom}\times\SI{10.1}{\angstrom}$~\cite{Afanasiev2021}.
		\textbf{b}~A $\textrm{NiPS}_3$ flake with 160\,nm thickness and lateral width around \SI{50}{\um} on top of a $\textrm{SiO}_2/\textrm{Si}$ substrate shows narrow PL emission (full-width at half maximum $\lesssim\SI{600}{\micro\eV}$) with strong linear polarization reaching 83\,\%. Spectra represent minimum (grey) and maximum (blue) emission. PL intensity versus polarizer angle $\theta$, plotted in the inset, is well described by $I(\theta)=I_0\sin^2(\theta)$, where $\theta$ denotes the analyzer angle with respect to the minimum, $I_0$, of the PL intensity. Blue and grey spectra respectively represent $90\deg$ and  $0\deg$ analyzer angles. 
		\textbf{c}~Optical absorption derived from the linear reflectance spectrum reveals three resonances, $X_1$, $X_2$, and $X_3$, below the charge transfer gap. 
		\label{fig:1}}
\end{figure*}

Within the class of vdW magnets, the family of transition metal thiophosphates ($\textrm{MPX}_3$, with $M$ being a transition metal and $X$ a chalcogen) realizes different types of antiferromagnetism, such as Néel, stripy, and zig-zag phases~\cite{Sivadas2015}.
$\textrm{NiPS}_3$ is a correlated insulator that hosts zig-zag chains of ferromagnetically oriented Ni spins, which align antiferromagnetically inside each layer below the Néel temperature $T_N=\SI{155}{\K}$ (see \cref{fig:1}a)~\cite{Kim2018}.

Our low-temperature photoluminescence (PL) measurements of $\textrm{NiPS}_3$ crystals exfoliated onto standard $\textrm{SiO}_2/\textrm{Si}$ substrates (see \cref{fig:1}b) show a narrow exciton emission peak at 1.476\,eV, in agreement with recent studies~\cite{Kang2020,Hwangbo2021}. 
Due to their exceptional coherence, these excitons exhibit spectral widths as low as \SI{350}{\micro\eV} in high-resolution PL experiments~\cite{Hwangbo2021}. 
Our theoretical analysis of strongly correlated electronic states in $\textrm{NiPS}_3$ based on an extended Hubbard model (cf. Supplementary Section 1A) reveals excitonic transitions between S \textit{p}-orbitals with zero net-magnetization and spin-polarized, long-range ordered Ni \textit{d}-orbitals (cf. red ellipses in \cref{fig:1}a).
As a result, these excitons intrinsically couple to the magnetic order of Ni spins.
Their highly anisotropic dipole moment predicted by our model (Supplementary Section 1C) is in excellent agreement with the large degree of linearly polarized PL emission (>80\,\%) observed in our experiments (cf. inset of \cref{fig:1}b).  

The optical absorption spectrum derived from the linear reflectance contrast in \cref{fig:1}c shows three distinct resonances close to the excitonic transition observed in PL spectra.   
The first resonance, $X_1$, coincides with the PL emission at $1.476$\,eV and is therefore ascribed to the absorption of these spin-correlated excitons~\cite{Kang2020}. 
Another absorption peak, $X_2$, at $1.479$\,eV, as well as a much broader feature, $X_3$, at 1.498\,eV, were also reported in ref.~\cite{Kang2020}. 
Their line-shape and temperature dependence indicate that, like $X_1$, they are coupled to the long-range magnetic order and may be related to magnon modes~\cite{Gnatchenko2011,Kudlacik2020,Kang2020}.
Indeed, photo-excitation of these resonances with ultrashort laser pulses has recently led to the generation of coherent magnon oscillations~\cite{Belvin2021}.

As the temperature approaches $T_N$, all the excitonic features discussed above vanish (see Fig.~S3, S6 and S10).
This strong correlation between excitons and antiferromagnetism is also apparent in an external in-plane magnetic field, which simultaneously rotates the magnetic axis and the emitted photon polarization~\cite{Wang2021}. 
In its antiferromagnetic state, $\textrm{NiPS}_3$ thus presents an excellent opportunity for exploring strong light-matter coupling with spin-correlated excitons.

\begin{figure*}
	\includegraphics[width=0.7\linewidth]{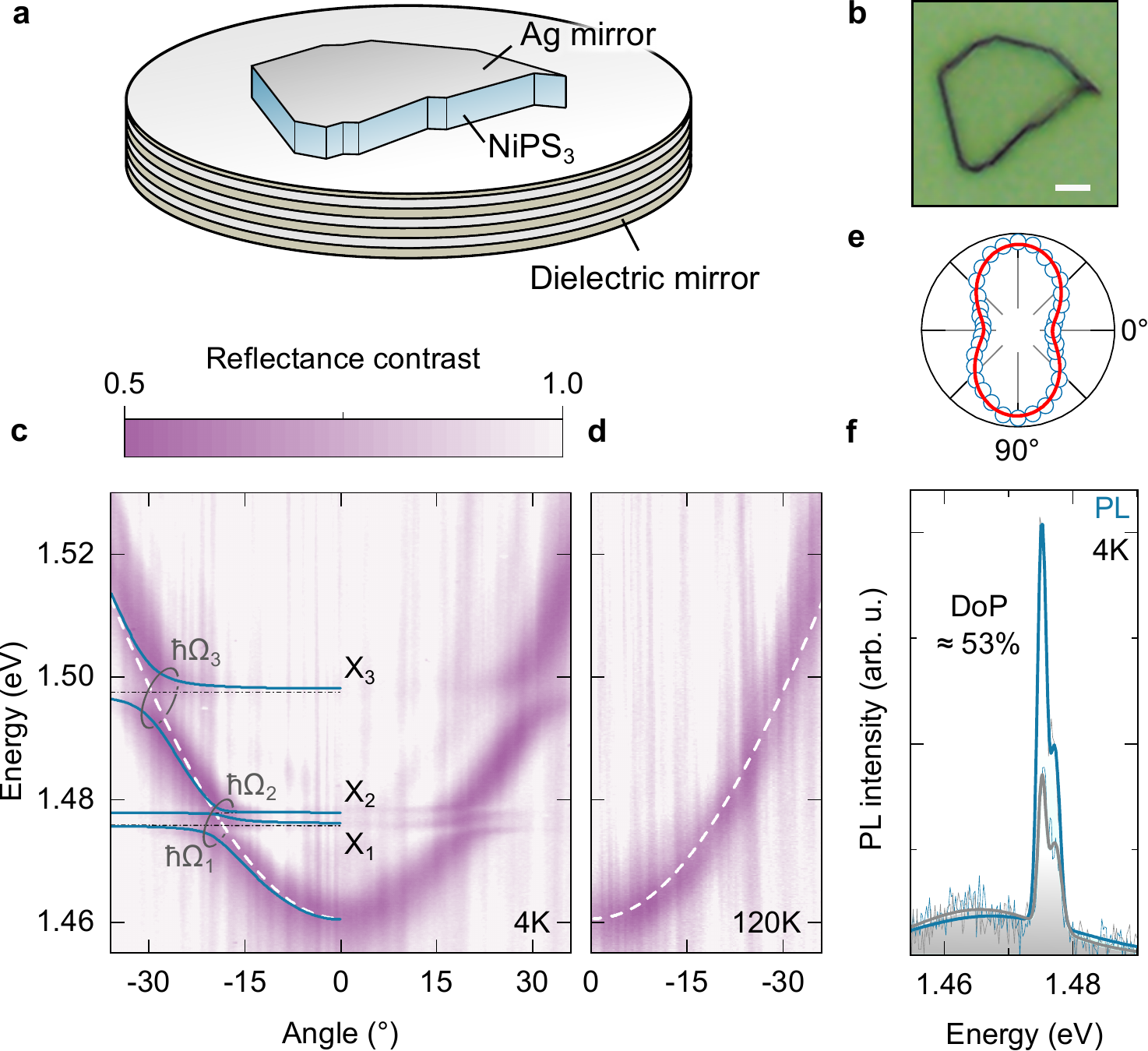}
	\caption{\textbf{Strong light-matter coupling in $\textrm{NiPS}_3$ microcavities.}
	\textbf{a}~$\textrm{NiPS}_3$ crystals enclosed by a dielectric bottom mirror and a 35\,nm-thin top silver layer form a microcavity.  
	\textbf{b}~Optical microscope image of a single $\textrm{NiPS}_3$ crystal inside the cavity. Scale bar is 5\,$\mu$m.  
	\textbf{c}~Angle-resolved optical reflection contrast map at 4\,K of the same crystal plotted together with a coupled oscillator model of a single cavity mode and absorption resonances $X_1$, $X_2$ and $X_3$ for negative angles (cf. also reflectance map and profile analysis in Fig.~S7).
	Anti-crossings at each intersection between the cavity mode (white dashed line) and an excitonic resonance (black dash-dotted lines) are reproduced with good agreement between the model and the experimental data. The resulting multiple branches of the polariton dispersion are depicted as blue solid lines. The Rabi splittings indicated by gray circles are $\hbar\Omega_1=\SI{4}{\meV}$, $\hbar\Omega_2=\SI{2}{\meV}$, and $\hbar\Omega_3=\SI{10}{\meV}$, as obtained from fits. 
	\textbf{d}~Anti-crossing features are almost absent in the reflectance contrast map recorded at 120\,K. 
	\textbf{e}~Integrated PL intensity of $X_1$ as a function of analyzer angle. 
	\textbf{f}~Angle-integrated PL emission shows strong linear polarization of more than 50\,\% and pronounced peaks at the $X_1$ and $X_2$ absorption resonances. Blue and grey spectra represent $90\deg$ and $0\deg$ analyzer angles, respectively. 
	\label{fig:2}}
\end{figure*}

To realize strong coupling between these exotic quasiparticles and microcavity photons, we transfer $\textrm{NiPS}_3$ crystals onto dielectric bottom mirrors prior to evaporating a 35\,nm-thin top silver mirror (see Methods and \cref{fig:2}a).
The thickness of each individual crystal directly determines the overlap with the photon cavity field.
About one out of four $\textrm{NiPS}_3$ crystals exhibits a cavity resonance suitable for the observation of strong coupling in our experiments (cf. Fig.~S4).  
\Cref{fig:2}b displays an optical image of a $\textrm{NiPS}_3$ microcavity flake with lateral dimensions around \SI{15}{\um} and a thickness of 170\,nm determined by atomic force microscopy. 

\Cref{fig:2}c shows an angle-resolved optical reflectance contrast map measured at $T$=4\,K. 
The photonic mode has multiple anti-crossing features that are characteristic of polariton formation via strong coupling of cavity photons and excitons. 
As indicated by the splittings, each absorption resonance ($X_1, X_2, X_3$) strongly couples to the optical cavity mode. 
We fitted the new modes using a coupled oscillator model (cf. Supplementary Section 6) and determined Rabi splitting energies $\hbar\Omega$ comparable to the values observed in standard band semiconductors~\cite{Tartakovskii2000}.
However, the anti-crossing signature of strong coupling vanishes upon approaching $T_N$ (cf. \cref{fig:2}d and Fig.~S10), indicating a relation between the polaritons and long-range magnetic order. 

First, we study the cavity PL emission obtained under non-resonant laser excitation at $T$=4\,K. 
Angle-integrated spectra comprise a main peak at 1.476\,eV, a pronounced high-energy shoulder at 1.478\,eV and a broad emission band around 1.46\,eV (cf. \cref{fig:2}f).
The linear polarization observed in \cref{fig:2}e shows that the antiferromagnetic order of $\textrm{NiPS}_3$ is preserved in our strongly coupled cavity sample. 
Like in the bare flakes on $\textrm{SiO}_2/\textrm{Si}$, the degree of polarization varies from sample to sample.

\begin{figure*}
	\includegraphics[width=0.7\linewidth]{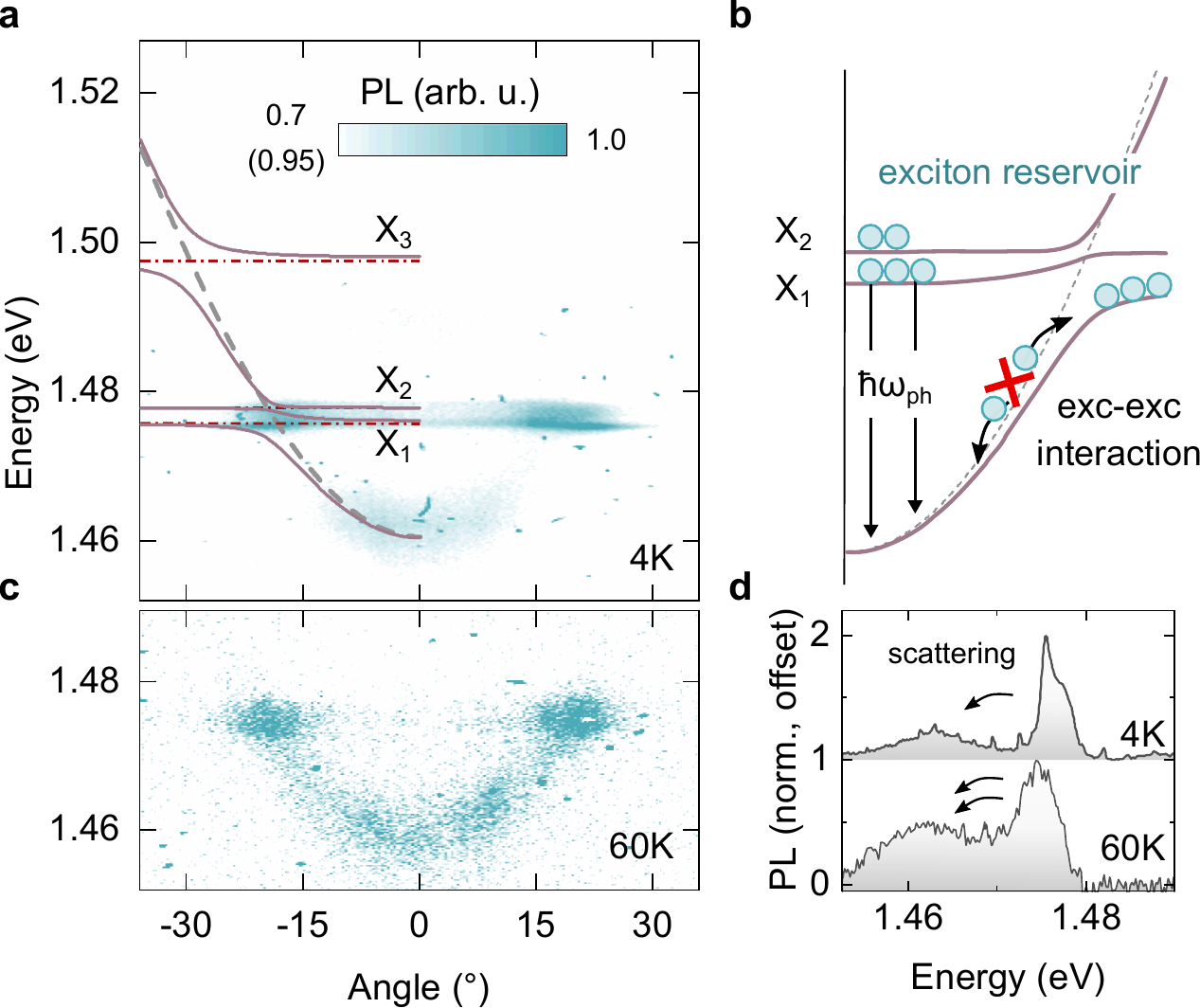}
	\caption{\textbf{Pronounced bottleneck of polariton relaxation.}
		\textbf{a}~Angle-resolved PL emission recorded at 4\,K superimposed by the coupled oscillator model shown in \cref{fig:2}c. 
		\textbf{b}~Schematic illustration of polariton relaxation and the bottleneck effect: Excitons from the reservoir scatter with phonons to weakly populate the lowest polariton branch, while efficient scattering due to long-range exciton-exciton interactions is suppressed.
		\textbf{c}~Cavity PL emission at 60\,K. Intensity-scale is indicated in \textbf{a} by the number given in brackets.
		\textbf{d}~Comparison of angle-integrated normalized PL spectra at 4\,K and 60\,K  (intensity spikes were removed).
		\label{fig:3}}
\end{figure*}

We gain valuable insight into the nature of the resulting exciton-polaritons from angle-resolved emission maps.
Unlike in most exciton-polariton systems~\cite{Deng2010,Virgili2011}, the strongest PL signal of our $\textrm{NiPS}_3$ cavity is observed at energies close to the $X_1$ and $X_2$ exciton resonances (see \cref{fig:3}a,b) and only a small fraction of the total polariton emission occurs from the lowest energy states of the polariton dispersion. 
This behavior indicates the presence of a pronounced polariton relaxation bottleneck towards the bottom of the dispersion at 1.46\,eV, and thus weak polariton scattering processes and interactions.
Despite matching with phonon energies found in recent Raman studies ($\Delta E\approx\SIrange[range-units = single,range-phrase = -]{10}{20}{\meV}$)~\cite{Kim2019}, exciton scattering rates are not sufficient to populate these low-energy polariton states during the short polariton lifetime~\cite{Zasedatelev2019}. 

Even in the absence of efficient polariton relaxation via phonons, long-range exciton-exciton exchange interactions typically overcome the relaxation bottleneck and facilitate majority population of low-lying polariton states~\cite{Deng2010}. 
However, the observed weak polariton relaxation indicates that these interactions are greatly suppressed in $\textrm{NiPS}_3$. 
As shown in \cref{fig:3}c,d, only the enhanced phonon scattering at elevated temperatures is able to reduce the relaxation bottleneck and thus increases the fraction of PL observed from low-energy polariton states around $\SI{1.46}{\eV}$.

\begin{figure*}
	\includegraphics[width=0.8\linewidth]{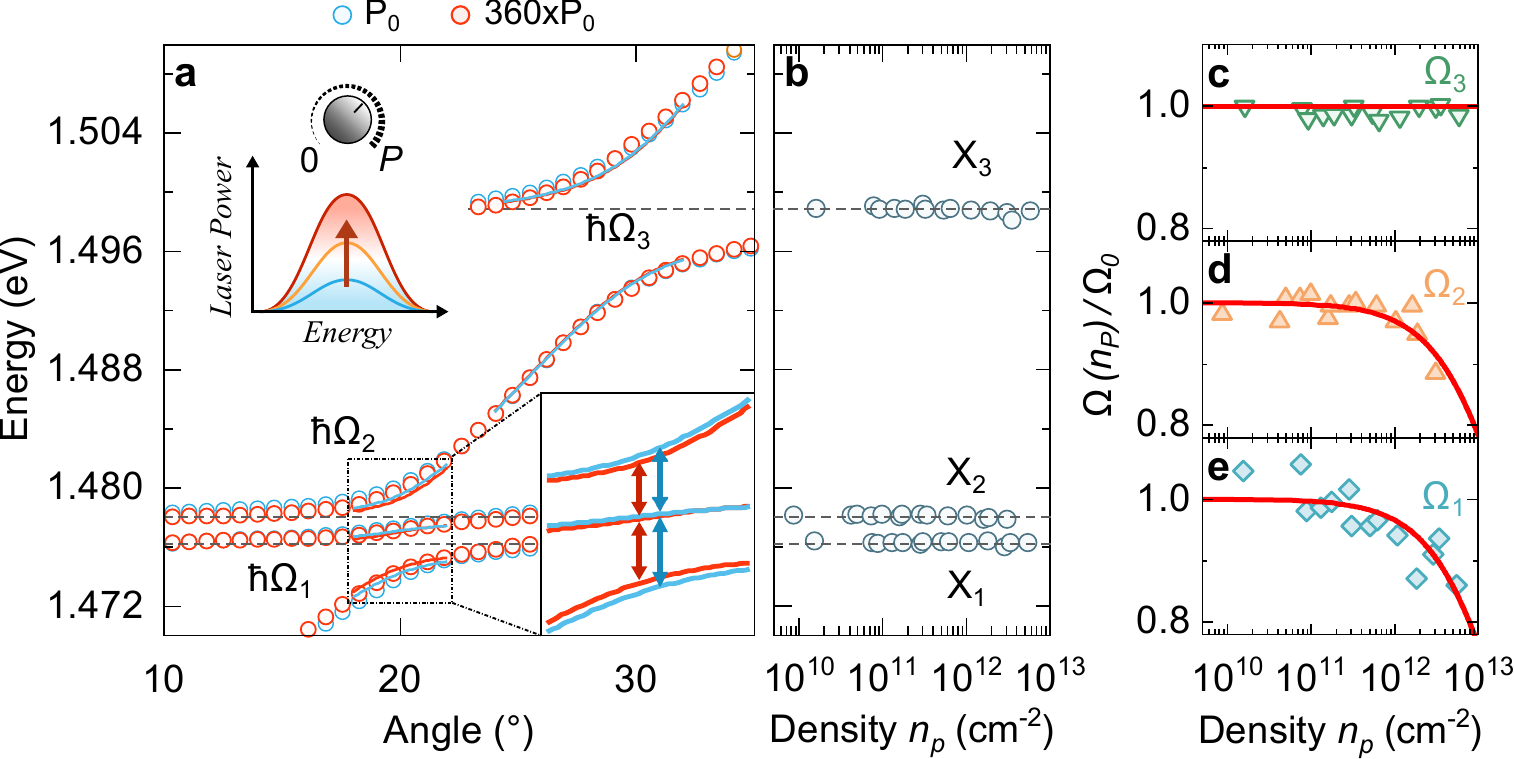}
	\caption{\textbf{Polariton non-linearities under increasing density.}
		\textbf{a}~Polariton dispersion for the lowest ($P_0$) and highest ($\approx 360 P_0$) excitation power of the incident broadband laser pulses. Inset shows a magnified view of the fitted oscillator model for both powers. 
		\textbf{b}~Density-dependent exciton energies directly determined from absorption minima in line-cuts at \SI{0}{\degree} ($X_1$ and $X_2$) and at \SI{20}{\degree} ($X_3$) with high exciton fraction. 
		\textbf{c}-\textbf{e}~Reduction of Rabi splitting $\Omega(n_p)$ relative to the Rabi splitting $\Omega_0$ obtained at the lowest excitation power. Red solid lines represent a fit based on \cref{Eq: 1} to determine the saturation density.
		\label{fig:4}}
\end{figure*}

We also perform polariton excitation spectroscopy -- a powerful technique highly susceptible to non-linear optical phenomena~\cite{Deng2010,Gu2021,Zhang2021} -- to further study the interactions of spin-correlated excitons (see Methods). 
As shown in \cref{fig:4}a, tuning the power density of the incident broadband laser pulses allows us to measure the microcavity dispersion under increasing excitation density. 
Since the excitonic fraction is dispersive, exciton energies can be obtained from absorption minima in line-cuts taken at angles at which the exciton fraction is close to unity (see Fig.~S9). 
Additionally, Rabi splitting energies can be determined from fits to the dispersion in the anti-crossing region.

Due to repulsive long-range interactions and the Pauli exclusion principle, polaritons formed by delocalized excitons in band semiconductors typically respond to increasing density by shifting towards higher energies~\cite{Gu2021,Zhang2021}.
However, for the three exciton resonances plotted in \cref{fig:4}b, no significant shifts are noticeable despite the high polariton densities ($n_P\approx\SI{5e12}{\cm^{-2}}$) reached in our experiment. 
Very similar behavior has recently been observed for localized moiré excitons in 2D vdW heterostructures~\cite{Zhang2021}.
In excellent agreement with our microscopic model presented in \cref{fig:1}a and the conclusions drawn from \cref{fig:3}, excitons in $\textrm{NiPS}_3$ are thus characterized by the suppressed long-range interactions of strongly localized excitons.

Despite the absence of intrinsic dispersive exciton nonlinearities, strong coupling introduces a collective nonlinear response in the electronic dispersion of our $\textrm{NiPS}_3$ microcavity.
Saturation of optical absorption modifies the dispersion due to a gradual reduction of Rabi splitting energies $\Omega(n_p)$ under increasing polariton density $n_p$ (cf. \cref{fig:4}c-e). 
While the negatively detuned cavity analyzed in \cref{fig:4} indicates no noticeable saturation for $\Omega_{3}$, higher total polariton density reached in a near zero-detuned cavity reveals its saturation at high powers (cf. Fig.~S12 and S13).
The large saturation density ($n_s$) of $X_1$ and $X_2$ exceeding \SI{e13}{\cm^{-2}} indicates that excitons in $\textrm{NiPS}_3$ are strongly localized and tightly bound by attractive Coulomb forces, despite the pronounced electrostatic screening effects inherent to bulk crystals. 
As expected from our microscopic model presented in \cref{fig:1}a, the average Bohr radius of $X_1$ and $X_2$ excitons in $\textrm{NiPS}_3$ estimated from their optical saturation ($a_B=\SI{6\pm1}{\angstrom}$, see Supplementary Section 10) is comparable to the order of the size of the unit cell~\cite{Afanasiev2021}. 
This conclusion is also in excellent agreement with recent first-principles calculations~\cite{Birowska2021} and the large effective masses predicted in ref.~\cite{Lane2020} resulting in Bohr radii well below 1\,nm.
The apparent inconsistency between the large exciton binding energies (several hundred meV) and the fact that spin-correlated excitons vanish around $T=T_N$ highlights the distinct origin of these magnetically coupled excitations in $\textrm{NiPS}_3$ from excitons in conventional band semiconductors.
Despite their exceptionally small Bohr radii, the modest Rabi splitting energies observed in our microcavities further demonstrate the presence of a relatively small exciton oscillator strength in $\textrm{NiPS}_3$ (see Supplementary Section S9). 
Finally, we emphasize that pronounced electronic correlations in $\textrm{NiPS}_3$~\cite{Kim2018} are responsible for the strong exciton localization supported by the results in \cref{fig:3,fig:4}.

In summary, we demonstrate strong coupling between microcavity photons and spin-correlated excitons in the vdW magnetic insulator $\textrm{NiPS}_3$. 
Hybridization leads to a new type of quasiparticle which inherits signatures of excitons, photons, and spins.
A detailed experimental and theoretical analysis of these polaritons demonstrates the absence of typical long-range exciton-exciton interactions to result from the strong localization of tightly bound excitons. 
Future studies could explore the potential of spin-correlated polaritons and their coupling to magnons for optical control of the high-speed information processing demonstrated in antiferromagnets~\cite{Jungwirth2018}.
Furthermore, with the ability of spin-correlated excitons to strongly couple to a microcavity photon field emerges the unique opportunity of realizing polariton Bose-Einstein condensates interacting with magnetically ordered spin systems.
Finally, our experiments on correlated vdW antiferromagnets represent a first step towards the cavity-induced optical engineering of strongly correlated quantum materials.



%
\textbf{Acknowledgments:}
We thank Andrew Millis, Alexander Punnoose and Xiaodong Xu for useful discussions. 

Work at CUNY was supported through the NSF QII TAQS 1936276 (V.M.M.) and the Army Research Office MURI grant: W911NF-17-1-0312 (V.M.M. \& A.H.M.). 
E.B. acknowledges support from the Robert A. Welch Foundation (grant F-2092-20220331). 
A.K. acknowledges support from a Graduate School Continuing Fellowship at the University of Texas at Austin. 
F.D. was funded by the Deutsche Forschungsgemeinschaft (DFG, German Research Foundation) through Projektnummer 451072703.

\textbf{Author contributions:}
V.M.M., E.B., F.D. and R.B. conceived the experimental idea and interpreted the results together with A.H.M. and A.K., F.D. and R.B. performed the experiments and conducted the data analysis with assistance from B.D., F.D. wrote the manuscript with input from all authors and V.M.M. supervised the project.

\textbf{Competing interests:} 
The authors declare that they have no competing interests.

\textbf{Data availability:}
The data sets generated during and/or analysed during the current study are available from the corresponding authors on reasonable request.


\bibliography{Manuscript_NiPS3-SC}
\nocite{ellis2008,negele1982,ring2004,kumar2021,Schmitt1985}

\section*{Methods}
\subsection*{Sample fabrication and characterization}
For microcavity fabrication, highly reflective Bragg mirrors (from 1.39\,eV to 1.62\,eV) were grown by plasma-enhanced chemical vapor deposition of 30 pairs of Silicon nitride/Silicon dioxide layers on Silicon wafers. 
Commercial $\textrm{NiPS}_3$ bulk crystals (acquired from 2D Semiconductors) were thinned down in multiple cycles of stick-and-release on blue tape (PVC tape 224PR, Nitto) and transfered onto a polydimethylsiloxane (PDMS, AD series, Gel-Pak) film. 
After moderately pressing the PDMS film onto a Bragg mirror (or a $\textrm{SiO}_2/\textrm{Si}$ substrate, oxide thickness \SI{285}{\nm}) and swiftly releasing it, the majority of crystals that were transfered showed a thickness of a few tens to hundreds of nanometers. 
Subsequent to $\textrm{NiPS}_3$ transfer, a 35\,nm-thin layer of silver was deposited by electron-beam physical vapor deposition to form a top mirror, resulting in measured quality factors of around 100. 
Within 24 hours after transfer, the samples were cooled down to $T$=4\,K and kept inside the high-vacuum chamber of a dry-flow cryostat (Montana Systems). 

\subsection{Optical spectroscopy}

Reflectance spectra and maps were obtained from the attenuated output of a spectrally broadband tungsten-halogen lamp (250W LSH-T250, Horiba), spatially filtered, and focused to a spot size of $\SI{2.0}{\um}$ by the $\SI{100}{\times}$ microscope objective (NA=0.66) mounted inside the cryostat. 
The reflected signal was analyzed by a linear polarizer along the axis of maximum PL intensity. 
Angle-integrated and angle-resolved measurements were recorded by respectively focusing either real-space images or the back-focal plane of the objective onto a spectrometer connected to a charge-coupled device (CCD). 
Polariton dispersions were fitted by varying the vacuum Rabi splitting and the exciton–photon detuning while keeping the exciton energy and the effective refractive index of $\textrm{NiPS}_3$ constant.

For the PL experiments, unless specified otherwise, a continuous-wave laser with 2.33\,eV energy and $\bar{P}=\SI{120}{\kW/cm^2}$ average output power was focused onto the sample to a spot size of $\SI{1.0}{\um}$ by the same objective used for reflectance spectroscopy. 
The collected PL signal was directed towards the spectrometer and spectrally filtered to remove the laser emission. 
For polarization-resolved measurements, stationary linearly polarized laser excitation was used to obtained PL emission, which was subsequently passed through a rotating half-waveplate and a polarizer fixed along the direction parallel to the entrance slit of the spectrometer.
The degree of linear polarization is thus defined as $\rho=(I_{\perp}-I_{\parallel})/(I_{\perp}+I_{\parallel})$, where $I_{\perp}$ ($I_{\parallel}$) denotes the PL intensity when the emission polarization is analyzed perpendicular (parallel) to the antiferromagnetic spin orientation~\cite{Wang2021}. 

\subsection{Nonlinear polariton spectroscopy}

For nonlinear spectroscopy, spectrally broad pulses from a super-continuum laser (NKT Photonics, SuperK Extreme) with a repetition rate of 78\,MHz and a pulse width of 20\,ps were filtered by a bandpass filter from 1.41\,eV to 1.52\,eV to minimize absorption at higher energies and the resulting heating effects. 
The time-averaged laser power of the spectrally filtered pulses was increased from 0.07\,mW to 25.3\,mW, while the reflected signal was attenuated by optical density filters before entering the detector.
Polariton branches were monitored for different input powers by mapping the numerically determined reflectance minima at different angles.
The resulting dispersion plots were then fitted in a small angle range around the anti-crossing region for each laser power with a coupled oscillator model to determine the Rabi splitting energies. 

Under increasing laser-induced polariton density $n_p$, saturation of optical absorption modifies the dispersion due to a gradual reduction of Rabi splitting energies $\Omega(n_p)$. 
The decrease with respect to the Rabi splitting obtained at the lowest density, $\Omega_{0}$, is given by~\cite{Zhang2021}
\begin{equation}
	\frac{\Omega(n_p)}{\Omega_{0}}=\frac{1}{\sqrt{1+\frac{n_p}{n_s}}}\,. \label{Eq: 1}
\end{equation}
Fitting \cref{Eq: 1} to the experimental data yields a saturation density $n_s$ for each set of polariton branches (see Supplementary Section S9).

\end{document}